\documentclass[superscriptaddress,groupedaddress,nofootnoteinbib,11pt]{article}  

\usepackage{graphicx}  
\usepackage{dcolumn}   
\usepackage{bm}        
\usepackage{jcappub}

\def\be{\begin{equation}}
\def\ee{\end{equation}}
\def\ba{\begin{eqnarray}}
\def\ea{\end{eqnarray}}

\def\d{\mathrm{d}}
\def\p{{\cal P}}

\def\L*{{\cal L}_*}
\def\L{\mathcal{L}}
\def\({\left(}
\def\){\right)}

\def\p{\partial}

\def\p{\partial}

\def\<{\langle}
\def\>{\rangle}

\def\cs2{c_{s}^{2}}

 \def\p{\partial}

 \def\bea  {\begin{eqnarray}}   \def\eea  {\end{eqnarray}}
 \def\bean {\begin{eqnarray*}}  \def\eean {\end{eqnarray*}}

\begin{document}

\title{Trispectrum from Co-dimension 2(n) Galileons}

\author{Matteo Fasiello}
\affiliation{CERCA/Department of Physics, Case Western Reserve University, 10900 Euclid Ave, Cleveland, OH 44106, USA}

\abstract{A generalized theory  of multi-field Galileons has been recently put forward. This model stems from the ongoing effort to embed generic Galileon theories within brane constructions. Such an approach has proved very useful in connecting interesting and essential features of these theories with geometric properties of the branes embedding.  We investigate the  cosmological implications of a very restrictive multi-field Galileon theory whose leading interaction is solely quartic in the scalar field $\pi$ and  lends itself  nicely to an interesting cosmology. The bispectrum is characterized by a naturally small amplitude ($f_{NL}\lesssim 1$) and an equilateral shape-function. The trispectrum of curvature fluctuations has features which are quite distinctive with respect to their $P(X,\phi)$ counterpart.\\
We also show that, despite an absent cubic Lagrangian in the full theory, non-Gaussianities in this model cannot produce the combination of a small bispectrum alongside with a large trispectrum. We further expand on this point to draw a lesson on what having a symmetry in the full background independent theory entails at the level of fluctuations and vice-versa.}


\maketitle


\section{Introduction}
Quantum field theories endowed with Galileon symmetry are ubiquitous in recent literature. They were first seen at work in the context of brane constructions, specifically as a  limit of the DGP \cite{DGP} model and have been investigated ever since. One can study their higher dimensional origin or be content with an effective 4D theory, obtained integrating out the bulk, which inherits most interesting properties. It is this last route that we will follow here but we refer the interested reader to the literature for the former perspective: \cite{DGP,reunited,cod2n} .

Among the most compelling properties which originate from the galilean symmetry characterizing these theories are second order equations of motion (from a theory with higher order derivatives) and non-renormalization theorems that protect the coefficients of Galileon terms from large renormalization.  The first property translates into a theory with a well defined Cauchy problem which is possibly unitary in the quantum regime and the second into a model which is \textit{predictive} as, under suitable conditions, the number of terms which describe the theory stays finite and their coefficients are only subject to small corrections.

It would then be intriguing to see if  any such theory can accommodate for an inflationary phase.
Recently, an interesting concrete realization of such a model, \textit{Galileon inflation}, has been proposed in \cite{Andrew1}: it  represents one of the very few radiatively stable and properly predictive models of inflation to date. This model shares with DBI \cite{DBI} inflation essential properties such as being unitary and having symmetries at one's disposal which protect leading operators from large quantum corrections.

DBI inflation provides us with a very compelling inflationary mechanism embedded in a UV finite theory such as  string theory. However, at least for its single-field-disguise, DBI predictions are quite close to be ruled out by data; the Planck mission is likely to narrow the bounds on cosmological observables which already look unfavorable for this model (not so for multi-field models, see e.g. \cite{Langlois:2008wt,Langlois:2008qf,Mizuno:2009mv,Mizuno:2009cv}).

The results of \cite{Andrew1} are compatible with current observational constraints: the bispectrum of curvature perturbation has been studied and predictions for the observable $f_{NL}$ have been extracted. The calculation of the corresponding trispectrum has been performed in \cite{tris1} (see \cite{Gao,tris0} for different but related approaches). 
As per usual, given a model for primordial perturbations, one would like to obtain the corresponding predictions on all available observables, chief among which  are non-Gaussianities (NG) \cite{Nic}. Ideally, NG of a given model are within experimental values and present distinctive features in the form of the so-called higher order correlators \textit{amplitudes} and \textit{shape-functions}.

Considering the overwhelming amount of inflationary models which are in agreement with observations and predict more or less the same scenario for CMB observables, one, while waiting for further data,  might want to look for some guiding principles in the search for the most compelling inflationary theories. 

Below we elect the stability of the theory as paramount and introduce a stable inflationary model which has quite distinctive signatures at the level of the trispectrum of curvature fluctuations. Our model is based on the co-dimension 2 (quite in general, it would suffice to have co-dimension 2n) Galileon theories introduced in \cite{cod2n} and the way we implement an inflationary mechanism is in close correspondence to what is done in \cite{Andrew1}. 

The authors of \cite{cod2n} show how an even co-dimension forces an additional symmetry $(SO(N))$ on the Galileon theory which propagates down to the 4D effective action and results in a Lagrangian with a quadratic and quartic piece only. We report here on the bispectrum of curvature fluctuations generated by such model and study in detail the trispectrum. We will show how the latter non-Gaussian observable is small and easily within constraints  provided by Planck \cite{Ade:2013ydc} (the same is known to be true for the bispectrum \cite{Andrew1}). The various shape-functions contributing to the trispectrum have distinctive features we shall illustrate in detail.   

We also offer some comments on the fact that,  when perturbing around a de Sitter background, the third order action for perturbations is not parametrically suppressed as one could naively have thought, this despite a null $\mathcal{L}_3$ in the full theory. A small-bispectrum---large-trispectrum scenario cannot therefore be implemented.

The paper is organized as follows: in \textit{Section 2} we review the model first introduced in \cite{cod2n}. In \textit{Section 3} we show how such a model can implement an inflationary phase. \textit{Section 4} is devoted to the calculation of non-Gaussianities. We offer some comments in the \textit{ Discussion} section, then summarize our findings and possible future work in the \textit{Conclusions}.

\section{Review of the Model}
Here we describe in some detail  the model first discussed in \cite{cod2n}. The key feature that identifies this model is the additional (with respect to, say, \cite{reunited}) $SO(N)$ symmetry it endows to the Galileon fields. This symmetry emerges from the requirement that the theory behaves as its co-dimension 1 counterpart and it is therefore both, connected with the essential interesting properties of Galileon theories, and  extremely natural.

Following \cite{cod2n}, we will take below a relatively quick route to the galilean action of interest; for a more systematic higher dimensional derivation of the same theory, we refer the reader to \cite{cod2n}.\\
In flat space, upon requiring that a theory is\\

\noindent $\bullet$ second order in the equation of motion and\\

\noindent $\bullet$ endowed with galilean symmetry $\pi \rightarrow \pi + c +b_{\mu} x^{\mu}$, \\ 

one arrives \cite{Nicolis:2008in} to the following general formula: 
\bea
 \mathcal{L}_{n+1}= \frac{1}{(n-1)!} \sum_{p} (-1)^{p} \eta^{\mu_1 p(\nu_1)} \eta^{\mu_2 p(\nu_2)}...  \eta^{\mu_n p(\nu_n)} \left( \partial_{\mu_1} \pi \partial_{\nu_1} \pi \partial_{\mu_2} \partial_{\nu_2} \pi ... \partial_{\mu_n} \partial_{\nu_n} \pi  \right); \quad \qquad \label{flatsingle}
\eea
there are d non trivial terms (plus the tadpole) in d space-time dimensions, five in 4D. Systematics ways of deriving the corresponding equation of motion for $\pi$ and currents associated with the shift symmetry have been already explored in the literature, and we do not report the results here. What concerns us is the generalization of Eq.~(\ref{flatsingle}) to the multifield case:

\bea
 \mathcal{L}_{n+1}=\frac{S_{I_1 I_2..I_{n+1}}}{(n-1)!}   \sum_{p} (-1)^{p} \eta^{\mu_1 p(\nu_1)} \eta^{\mu_2 p(\nu_2)}...  \eta^{\mu_n p(\nu_n)} \left(  \pi^{I_{n+1}}\partial_{\mu_1} \partial_{\nu_1} \pi^{I_1} \partial_{\mu_2} \partial_{\nu_2} \pi^{I_2} ... \partial_{\mu_n} \partial_{\nu_n} \pi^{I_n}  \right),\nonumber \\ \label{flatmulti}
\eea
\noindent where $S_{I_1 I_2..} $ is a tensor symmetric in the field index $I$. It is easily verified that this multifield generalization preserves the galilean symmetry on each field $\pi^{I} \rightarrow \pi^{I} + c +b_{\mu} x^{\mu}$ and retains equations of motion free of the Ostrogradski \cite{Ostro, Chen:2012au, JLM, Woodard} instability. As we will see, both in the single and multifield case the Galileon symmetry can be thought of as deriving from symmetries which characterize a higher dimensional brane construction \cite{reunited}.

Consider indeed the case of a 3-brane in a higher dimensional bulk (5D for now). Assume flat bulk and use $X^{A}(x)$ to describe the embedding of the brane, where $A$ signals the bulk dimensionality and $x$ describes the brane coordinates. We want any action to be invariant under Poincare transformations of the bulk and gauge invariant under reparametrization of the brane:
\bea
\delta_P X^A= \omega^{A}_{B} X^B +\epsilon^A; \qquad \delta_g X^A = \xi^{\mu}(x)\p_{\mu}X^A
\eea
where $\epsilon^{a}$ describes translations in the bulk, $\omega$ Lorentz transformations and $\xi$ is the usual gauge parameter. One can fix a gauge, e.g. unitary gauge, at this stage:
\bea
X^{\mu}(x)=x^{\mu}, \quad X^5(x)=\pi(x) \qquad \Rightarrow \qquad \delta_P X^{\mu}= \omega^{\mu}_{\nu} x^{\nu}+ \omega^{\mu}_{5} \pi + \epsilon^{\mu},
\eea
and so it becomes clear that an additional input is needed to have a gauge fixed action invariant under the Poincare transformation. Upon choosing:
\bea
\xi^{\mu}= -\omega^{\mu}_{\nu}x^{\nu}-\omega^{\mu}_{5}\pi-\epsilon^{\mu} ,
\eea 
the combined action  $\delta_{P'}=\delta_{P}+\delta_{g}$ transformation is a symmetry the gauge fixed action. The crucial bit now is to see the effect of  this transformation on $\pi$ which, the notation gives it away, is going to be our Galileon. Indeed the total transformation on $\pi$ reads \cite{reunited}:

\bea
\delta_{P'}\pi =-\omega^{\mu}_{\nu}x^{\nu} \p_{\mu}\pi-\epsilon^{\mu}\p_{\mu}\pi +\omega^{5}_{\mu}x^{\mu}-\omega^{\mu}_{5}\pi \p_{\mu}\pi +\epsilon^{5} \label{singletransform}
\eea
The first two terms are the unbroken 4D Poincare transformations, the second two terms are the broken boosts and the last one corresponds to broken translations in the fifth direction; in other words what happens is $ISO(1,4) \rightarrow ISO(1,3)$ . To get a glimpse of the galilean symmetry consider the internal relativistic invariance under which $\pi$:
\bea
\delta_{P'}\pi= +\omega_{\mu}x^{\mu}-\omega^{\mu}\pi \p_{\mu}\pi +\epsilon,
\eea
where the index $^{5}$ has been omitted. In the non relativistic limit (small $\pi$) this is the usual $\pi\rightarrow \pi+c +b_{\mu}x^{\mu}$, but we now see how this symmetry is originating from the brane motion in the (flat) bulk. This results lends itself to a straightforward generalization. Suppose that the co-dimension is not 1 anymore, but generic.  The only relevant difference for us will be on the transformation for $\pi$, now $\pi^{I}$, which, in the gauge fixed action under the combined effect of  $\delta_{P'} \pi^{I}= \delta_{P} \pi^{I} +\delta_{g}\pi^{I}$, gives:

\bea
\delta_{P'}\pi^{I} =-\omega^{\mu}_{\nu}x^{\nu} \p_{\mu}\pi^{I}-\epsilon^{\mu}\p_{\mu}\pi^{I} +\omega^{I}_{\mu}x^{\mu}-\omega^{\mu}_{J}\pi^{J} \p_{\mu}\pi^{I} +\epsilon^{I}+\omega^{I}_{J}\pi^{J}. \label{multitransform}
\eea

Let us focus on the last term in the above expression; we do so because it is the only one which is not a straightforward generalization of Eq.~(\ref{singletransform}). Indeed $\omega^{A}_{B}$ is antisymmetric and so its $\omega^{5}_{5}$ entry vanishes in the single field (that is, co-dimension $1$) case. This last term is signalling us an $SO(N)$ symmetry in the dimensions traverse to the brane, therefore on top of the usual galilean invariance on each field, in the non relativistic limit we now count an additional $SO(N)$.

This symmetry has striking consequences on the galilean Lagrangian of the type in Eq.~(\ref{flatmulti}). Indeed, with $S_{I_1,I_2...}$ a symmetric tensor, as a matter of simple indices contractions, one realizes the only allowed actions contain an even number of fields, in 4D we have just two different contributions! These are the usual kinetic term  and a quartic term in the field $\pi$:

\bea
\mathcal{L}_2=\p_{\mu}\pi^{I}\p^{\mu}\pi^{I}, \,\,\,\,\,\,\qquad \qquad\qquad\qquad\qquad\qquad\qquad\qquad\qquad\qquad\qquad\qquad\qquad\qquad\qquad\qquad  \\ \mathcal{L}_4=\p_{\mu}\pi^{I}\p_{\nu}\pi^{I}\left( \p^{\mu}_{} \p^{}_{\rho}\pi^{J} \p^{\nu}_{} \p^{\rho}_{}\pi^{J}-\p^{\mu}_{} \p^{\nu}_{}\pi^{J} \Box \pi^{J} \right) +\frac{1}{2} \p_{\mu}\pi^{I}\p^{\mu}\pi^{I}\left( \Box \pi^{J} \Box \pi^{J} -  \p_{\nu} \p_{\rho}\pi^{J}\p^{\nu}\p^{\rho}\pi^{J} \right), \nonumber  \label{model}
\eea
\noindent where the indices $I,J$ are contracted with a Kronecker delta. Having reached this point, it is useful to remark that both the renormalization and (second order) equation of motion properties of this multifield theory (the latter can be seen basically by inspection) are left untouched. It is the galilean symmetry that protects these theories from large renormalization corrections and such a symmetry is still intact.

The additional $SO(N)$ symmetry is the news (see also \cite{Padilla:2010ir,Padilla:2010de}) and, as we will see in the next section, it can be put to good use in cosmological scenarios. We note here that perhaps the most elegant route to the Lagrangian above is the one that arrives at it by writing down the most generic action with galilean symmetry and internal relativistic invariance (plus the requirement of 2nd order e.o.m. that leads essentially to Lovelock invariants \&Co). We refer the curious reader to \cite{reunited,cod2n} for further reading on these aspects.

As mentioned, Galileon theories are ubiquitous in the literature and have already been employed to probe different scenarios: modified gravity, but also early universe cosmology. This is certainly the case of the work in \cite{Andrew1}, where single field Galileon theories where shown to be compatible with an inflationary phase and to become the leading contributions to the inflationary dynamics in a specific,  interesting, regime.

Assuming for a second that there is indeed such a possibility also for the case in question here, Eq.~(\ref{model}), it is clear why such a theory looks interesting: we are presented with, besides the usual kinetic term, a single quartic Lagrangian, one coupling, which presumably can lead the dynamics of fluctuations during an inflationary phase.

 If, upon switching on fluctuations, we were to find a sizable region of the parameter space where the third order action is suppressed (such an occurrence is clear-cut fine-tuning in a scenario with $\mathcal{L}_3\not=0$ and no well justified symmetries), we would have landed on a \textit{stable}, \textit{predictive}, theory with a  naturally large four-point function and a parametrically small three-point function, that is $f_{NL}$. This is certainly interesting in light of the upcoming analysis of Planck results (see also \cite{Senatore:2010jy,Izumi:2010wm}).

\section{Background Analysis and Inflating Solution}
We will show below that our model supports an inflationary phase and that, thanks to the inherited $SO(N)$ symmetry,  it  can be essentially treated as a single field model for the purposes of this work.
In order for us to be able to trust the inflationary theory inherits all the desirable properties described above we will proceed in steps \cite{Andrew1}. \\

\noindent - We first consider an $M_{Pl}\rightarrow \infty$ dS limit and show that there the background is compatible with an inflationary solution and that the shift symmetry is broken, so that inflation can end.\\
- We then go on to show that the potential chosen, although it does break galilean invariance as well,  it is not renormalized \cite{Porrati:2004yi,Nicolis:2004qq,Endlich:2010zj}.\\

\noindent We start with:

\bea
\mathcal{L}_{T}=\sqrt{-g}\left( \frac{M_{Pl}^2}{2} R- \frac{\alpha_2}{2} \nabla_{\mu}\pi^I \nabla^{\mu}\pi^I
- \alpha_4 \mathcal{L}_4 (\pi^I) \right) 
\eea

\noindent One can make immediate use of the $SO(N)$ symmetry by choosing to specialize along the direction of one single field in all that follows.  This choice is consistent because it is protected by an exact symmetry the theory enjoys. It is a clear-cut, one-time-for-all, tuning of the initial conditions using $SO(N)$. As such, this leads to a completely consistent and predictive dynamics for the theory. One should think of this choice as limiting the phase space of the theory; it is the symmetry that assures us we will remain in this region over time. This tuning of the initial conditions is to be understood in contradistinction to a dynamical tuning of the parameters.

To be sure, by doing so we are (albeit consistently) partially betraying the fully multi-field nature of the model. In the analysis with more generic initial conditions one must take into account  distinctively multi-field aspects  (i.e. the conversion of isocurvature modes into adiabatic ones)   of the model. It is also true that they can depend on the specific  isocurvature-adiabatic conversion mechanism and our focus here is not on such matters. 
We have embarked in the fully multi-field treatment of the theory under scrutiny, and we refer the interested reader to further work of ours which is nearing completion \cite{multi}.

\vspace{.4cm}

\noindent It is instructive to consider the $M_{Pl}\rightarrow \infty$ limit obtained by keeping the background fixed to de Sitter and by adding a potential of the form $V=V_0-\lambda^3 |\vec{\pi}| $. The potential is needed to (eventually) break shift symmetry, essentially for a graceful exit, and we chose its form here such that it is simple and such that the renormalization properties are safeguarded (going beyond a mass term,  i.e. $V\sim \pi^n, n>2$ , is not possible precisely for this latter reason) \cite{Andrew1}.

Being in curved space, we must now of course have a prescription for the covariant Galileon theory. As it happens a naive $\p \rightarrow \nabla$ covariantization does not work, as second order e.o.m. would be lost.   On the other hand, a covariantization procedure has been worked out in \cite{Deffayet:2009wt,Deffayet:2009mn,Deffayet:2010zh} and that immediately applies to our $\mathcal{L}_4$. The action for the background field $\bar \pi(t)$ is:

\bea
S_b=\int a^3\Big[ {\frac{\alpha_2}{2}\dot{\bar{\pi}}}^2 +\frac{9}{2}\,\alpha_4 \frac{H^2}{\Lambda^6} \dot{\bar{\pi}}^4   \Big].
\eea
\noindent Following \cite{Andrew1} , we now point out several crucial ingredients:\\

- the shift symmetry of the potential is not broken in this limit as any piece linear in $\bar \pi$ (think of $\sqrt{-g}\,\lambda^3\, \bar \pi$) is a total derivative. We conclude that any breaking of the shift symmetry is $M_{Pl}$ suppressed.\\

- the equation of motion for ${\bar{\pi}}$ is compatible with the $\dot {\bar{\pi}} = const $ solution. The latter is crucial for the non-renormalization properties of the theory: indeed whenever one states that Galileon theories are not renormalized this is true provided higher order derivatives such as $\ddot\pi,  \dddot\pi$ are zero or kept small, otherwise other terms would emerge and our analysis could not be limited to terms such as the ones in Eq.~(\ref{flatmulti})

Precisely because in this limit the shift symmetry $\pi \rightarrow \pi+const$ is conserved (the potential is a total derivative and everything else counts at least one derivative per scalar) one can derive a current and identify the solution for $\bar \pi$ from the current conservation equation. Here we shall be content with the e.o.m. from the above action, which reads:
\bea
\Big[...\Big] \ddot{\bar{\pi}}+3H \dot{\bar{\pi}} \Big[\alpha_2  + 18 \alpha_4 \left(\frac{H }{\Lambda^3}\right) \dot{\bar{\pi}} \Big]=\frac{\lambda^3}{3 H}
\eea
\noindent so that $\dot{\bar{\pi}}=const$ is indeed a solution for $\bar \pi$ and therefore the renormalization properties, at least in this $M_{Pl} \rightarrow \infty$, dS-fixed limit, stay intact. This is not necessarily true later, when one accounts for gravity and the Galileon symmetry itself is broken, but in that regime the breaking will be parametrized by $\Lambda/M_{Pl}$, which we can safely assume to be small. 

In other words, what we have been after in this section can be summed up as showing that everything works out nicely in this limit and therefore, whatever happens in other regimes, will be at the very worst parametrized by $\Lambda/M_{Pl}$ ($H/{M_{Pl}}$ being the other possibility) and so should be under control. We have shown that the shift symmetry is preserved and argued that considering $\ddot {\bar\pi}=0$ and the specific choice of potential \cite{Andrew1}, also the non-renormalization theorems  still apply.

Having established that the Galileon terms are well behaved, we must find out in what regime they are relevant. Indeed, already at the background level, we see that
\bea
\mathcal{L}\sim \frac{\alpha_2}{2} \dot {\bar{\pi}} +\frac{9}{2}\,\alpha_4 Z^2 \frac{H^2}{\Lambda^6} \dot{\bar{\pi}}^2, 
\eea
where $Z=H \dot{\bar{\pi}}/\Lambda^3$. It might not be extremely suggestive in this form, but the less restrictive theory without $SO(N)$ symmetry where $\mathcal{L}_3, \mathcal{L}_5$ survive (that is, the explicit expression for Eq.~ \ref{flatmulti}) leaves no doubt that $Z$ is acting like a coupling constant for Galileon theories and therefore: as long as $Z\lesssim 1$ these theories have little new to say, but whenever  $Z$ is greater than unity, we depart from canonical inflation and Galileon non-linearities must be accounted for, they are indeed the leading contributions.

It is clear at this stage that Galileon non-linearities are the leading terms in the action just above in the $Z\gtrsim 1$ regime, not so much that they might be leading with respect to metric perturbations as well. In order to show that it is indeed the case, we need a quick detour into the effective field theory of inflation (more precisely ``of fluctuations around an FRW solution'') which was set up in \cite{Cheung} and further generalized in \cite{Senatore:2010wk} (for non-Gaussianities produced within this approach see also \cite{Senatore:2009gt,Bartolo:2010bj,Bartolo:2010di,Bartolo:2010im,Fasiello:2011fj}).
\subsubsection*{Relation to DBI-Galileon Models of Inflation}
We elucidated above the properties of this restricted set of Galilean interactions within the context of higher-dimensional constructions. Recently, it has been shown \cite{reunited} that both DBI \cite{DBI} and Galileon models can be seen as different limits of the same theory. Both these higher-derivative theories share crucial properties such as having a well-defined Cauchy problem (in other words, no troubling instabilities will arise). The connection between DBI and Galileons is a higher-dimensional symmetry \cite{reunited}. 

The cosmology of so-called DBI Galileon models has been investigated in    \cite{Mizuno:2010ag,RenauxPetel:2011dv, RenauxPetel:2011uk, Koyama:2013wma, Renaux-Petel:2013ppa}. More precisely, the model we have here can be seen as a ``non-relativistic" limit of the one in \cite{RenauxPetel:2011uk} \footnote{On top of taking the ``non-relativistic" limit, it must be said that here we restrict the phase space of the theory to that of a single field model via SO(N), whereas the authors of \cite{RenauxPetel:2011uk} present a fully multi-field analysis. The work in \cite{multi} would then be tantamount to studying the trispectrum for the``non-relativistic" limit of \cite{RenauxPetel:2011uk}.}. That is indeed the limit where Galileon interactions such as the one we study here emerge. On the other hand, our setup can also be viewed from a purely 4D perspective as a theory with Galileon terms endowed with specific symmetries, much as is the case for the Galileon inflation model of \cite{Andrew1}.

\subsubsection*{On the E.F.T.I. Approach and Why It Is Safe to Neglect Metric Fluctuations}
In \cite{Cheung} the authors assume they have a stable effective theory around an $FRW$ background and work out the most generic form that the perturbations can take. They first choose to work in a unitary gauge where the scalar mode $\delta\phi=0$ ($\delta \pi$ for us here) is ``eaten by the metric''. This choice breaks time-reparametrization invariance and therefore the most generic theory will have only space diffeomorphisms as a symmetry. These are in short the premises that lead to the following Lagrangian:
\bea
S = \int d^4 x \sqrt{-g} \; F(R_{\mu\nu\rho\sigma}, g^{00}, K_{\mu\nu},\nabla_\mu,t) .
\eea
\noindent
As the free $g^{00}$ indices and the presence of the (essentially 3D) extrinsic curvature $K_{\mu\nu}$ signal, the action is indeed only invariant under space diffs. The crucial step now is to reintroduce full space-time reparametrization employing the so called Stueckelberg trick: one promotes the parameter that describes time reparametrization to a field $\tilde \pi$ (not to be confused with our $\pi$) which now has a specific gauge transformation and reinstates full diffeomorphisms.
This entire procedure is ideally cast for a decoupling limit: we want the dynamics of $\tilde \pi$ to decouple from gravity. To this aim, it suffices to consider the terms in the action that mix $\delta g, \delta \tilde\pi$. After the field promotion, one typical term will have the following form
\bea
M_{\rm Pl}^{2} \dot H g^{00} \rightarrow M_{\rm Pl}^{2} \dot H ((1+\dot{\tilde\pi})^2 g^{00} +2(1+\dot{\tilde\pi})g^{0i}\partial_i {\tilde\pi}+ g^{ij}\partial_i {\tilde\pi} \partial_j {\tilde\pi}).
\eea
Focusing on the quadratic part of the first term, upon canonically normalizing both ($\pi_{c}=M_{\rm Pl}{\dot H}^{1/2}{\tilde\pi}$ and $ g^{00}_{c}=M_{\rm Pl}g^{00}$), one finds 
\bea
M_{\rm Pl}^{2} \dot H(  g^{00} + 2 \dot {\tilde\pi}  g^{00} + \dot{\tilde \pi}^2 g^{00} )=   {\dot{\tilde\pi}_{c}}^2 + 2 {\dot H}^{1/2}\dot{\tilde\pi}_{c} g^{00}_{c}  + \dot H g^{00}_{c}. 
\eea
It is clear that, the $\tilde \pi$-only term having more derivatives, it will lead over any mixed term for sufficiently high energy. In this specific example:
\bea
E_{mix} \gtrsim {\dot H}^{1/2},
\eea
where one should note that the value of $E_{mix}$ can change if the leading kinetic term for $\tilde\pi$ is different as is indeed the case in ghost inflation \cite{ArkaniHamed:2003uz,Senatore:2004rj}. This might surely appear to be a different approach than the one we take here, but it easy to intuitively spell out the dictionary: \\

\noindent - if we where to neglect metric perturbations, it is well known that our $\delta \pi$ would be linearly related to the curvature fluctuation $\zeta$.\\
\noindent - also, at first order in slow roll, neglecting $\delta g$, the following relation holds \cite{Cheung}:
\bea
\zeta=-H \tilde \pi,
\eea
that is, whenever the metric fluctuations are switched off, the perturbations of the scalar degree of freedom that drives inflation is, in both these languages, proportional to $\zeta$. It is then intuitively clear that there must exist for our $\delta \pi$, as much as it does exist for $\tilde \pi$, a regime in which metric fluctuations can be safely disregarded. It is in this regime that we choose to work below.

One might well ask at this stage why, if both the reasons and the procedure whereby we neglect the metric fluctuations are so clear-cut in the setup of \cite{Cheung} , we chose to work in our set up instead. The motivations are to be found in the previous section where the  background analysis was instrumental in tracking down the crucial properties of the theory as we have moved in steps towards the more realistic model which accounts for gravity. In order to do it, one needs to have a hold of the specific theory, a theory of its fluctuations is often not enough\footnote{It must also be said that, with reasonable additional hypotheses, a (possibly stable) and very interesting theory of ``Galileon inflation'' in a setup similar to \cite{Cheung} can indeed be constructed, see \cite{Creminelli1}. }. We return to a related point in the \textit{Discussion} section.

\section{Non-Gaussianities}
Having written down the theory and discussed its property at length, we move now to calculate observables such as higher order correlators of the curvature fluctuations $\zeta$. The current  bounds on the amplitude and profile of these quantities are, as well known, soon-to-be made more stringent by the analysis of the data provided by the Planck mission\footnote{Even after the first Planck data release, the most stringent bound one can find in the literature for the four-point function amplitude $\tau_{NL}$ generated by interactions such as the ones we study here is in a work by Fergusson et al. \cite{James}. There, the authors give a bound on $\tau_{NL}$ generated by interactions such as $(\dot{\zeta})^4$ (not identical, but qualitatively similar to the ones we have here) which reads $\tau^{eq.}_{NL}=(-3.11\pm 7.5)\times 10^6$. A fact that, as we shall see, suggests the predictions of the model studied here are well within observational bounds. Further constraints on similar interaction terms might soon be released by the Planck collaboration.} . 

Before presenting the actual trispectrum calculation, we pause here to note that, despite what one might naively expect, a null cubic Lagrangian in the full theory, does not easily lead to a small three-point function.
We can see this quickly by looking at the full Lagrangian and using an estimate of the main contributions to the  three(four)-point function. Consider  
\bea
			S =  \int \d^4 x \; \sqrt{-g} \, \Bigg[
				- \frac{\alpha_2}{2} (\nabla \pi)^2
				- \frac{\alpha_4}{\Lambda^6} ( \nabla \pi )^2 \Big\{
					(\Box \pi)^2 - (\nabla_\mu \nabla_\nu \pi)
					(\nabla^\mu \nabla^\nu \pi)
					- \frac{1}{4} R (\nabla \pi)^2
				\Big\} \Bigg],\nonumber 
\eea
\noindent the $\alpha_4$-regulated main bispectrum contribution will be of the type
\bea
S^{\alpha_4}_{3}\sim \frac{\alpha_4}{\Lambda^6} \dot{\bar {\pi}}\, \p \delta \pi (\Box \delta \pi)^2\sim \frac{\alpha_4} {\Lambda^6}\dot{\bar {\pi}} H^5 \delta\pi^3
\eea
use $\delta \pi \sim\dot{\bar {\pi}}\, \zeta/H $, to convert $\delta \pi$ into the proper observable, the curvature $\zeta$,
\bea
S^{\alpha_4}_{3}\sim \frac{\alpha_4} {\Lambda^6}\dot{\bar{\pi}} H^2\dot{\bar{\pi}}^3 \zeta^3=  \alpha_4\dot{\bar{\pi}}^2 Z^2 \zeta^3; \qquad {\rm where}\quad  Z\equiv \frac{H \dot{\bar{\pi}}}{\Lambda^3}. \label{bisest}
\eea
Proceeding in the same way for the trispectrum,
\bea
S^{\alpha_4}_{4}\sim \frac{\alpha_4}{\Lambda^6} \p \delta \pi \p \delta \pi (\Box \delta \pi)^2 \sim \frac{\alpha_4} {\Lambda^6} H^6 \delta \pi^4=\frac{\alpha_4} {\Lambda^6} \dot{\bar{\pi}}^2  \dot{\bar{\pi}}^2  H^2 \zeta^4= \alpha_4  \dot{\bar{\pi}}^2 Z^2 \zeta^4, \label{trisest}
\eea
\noindent where for this estimate we have assumed all along that the main contribution to correlators comes as usual from the around-the-horizon region where $\dot \pi\sim H \pi$ and $\p_i \pi \sim H/c_s \,\, \pi$. 

\noindent Summing up the estimate, we have
\bea
\mathcal{L}_3( \zeta^3)+ \mathcal{L}_4( \zeta^4)\sim \alpha_4 \dot{\bar{\pi}}^2 Z^2 (\zeta^3+\zeta^4) . \label{estimate}
\eea
\noindent The fact that the third and fourth order fluctuations have the same coefficient clearly means that there is no room for a small bispectrum vs large trispectrum finding. Furthermore, it also implies  that the relation between the corresponding amplitude parameters $f_{NL}$ and $\tau_{NL}$ is the standard one:  if this is the case, the value of $\tau_{NL}$ has to be about\footnote{This is also a model dependent statement, so ours here is to be considered a rough estimate.} five order of magnitude larger than $f_{NL}$ in order to be detectable. This is not what happens in the case of Eq.~(\ref{estimate}), as the $\tau_{NL}/f_{NL}$ ratio is much smaller, it stays too small for detection even if one is willing to allow as small a speed of sound \footnote{This value is close to current bounds on $c_s$ obtained by the corresponding bounds on $f_{NL}$ for quite generic effective field theory models which comprise a $c_s\ll 1$ dynamics (see also \cite{Senatore:2009gt}).  } as $c_s\sim 10^{-2}$ , which is far smaller than what is allowed in our specific setup.\\

With a possibly small $f_{NL}$ (compatible with zero) from Planck data in mind, one might have hoped that a stable, predictive and very restricted (by symmetries) model of inflation with a null cubic interaction in the full theory would be a good candidate for a naturally large four-point function whilst generating a small three-point correlator ($f_{NL}$). This would have granted a detectable $\tau_{NL}$.

For it to happen one  needs the (naively) additional background quantity $\dot{\bar{\pi}}$ which appears in (\ref{trisest}) to be small so that the three point function coefficient can be smaller than the four point function one. Upon comparing (\ref{bisest}),  (\ref{trisest})  we see that the relation $\delta \pi \sim\dot{\bar {\pi}}\, \zeta/H $ evens the score of background quantities thus making the two coefficients essentially of the same order.  We proceed with the full calculation below, returning to the lesson that can be drawn from this fact in the \textit{Discussion} section.

We now turn to a detailed analysis of fluctuation. These are obtained by expanding around an FRW background, to lowest order in slow roll, with $\pi(t,\vec x)= \bar{\pi}+\delta\pi(t, \vec x)$. We determine the wave function from the quadratic action:
\bea
S_2=\int a^3 \Bigg[\gamma_1 \dot\xi^2 -\gamma_2\frac{(\p_i \xi)^2}{a^2}\Bigg] ,
\eea
where 
\bea
\gamma_1=\frac{\alpha_2}{2}\dot{\bar{\pi}}^2+27 \alpha_4 Z^2 \dot{\bar{\pi}}^2 ; \qquad \gamma_2=\frac{\alpha_2}{2}\dot{\bar{\pi}}^2+ 13 \alpha_4 Z^2 \dot{\bar{\pi}}^2;  \qquad c_s \equiv \sqrt{\frac{\gamma_2}{\gamma_1}} ;\qquad  \delta \pi\equiv  \dot{\bar{\pi}} \xi\,\,. \qquad  
\eea
\noindent Notice that, short of indulging into serious fine tuning, the value of $c_s$ can safely be assumed to be in the  $1/2 <c_s\leq1$ interval. This will naturally have consequences for the value of the bispectrum amplitude, which cannot be very large.

\noindent The quadratic Lagrangian gives the usual dispersion relation and canonical wavefunction (we adopt here the Bunch-Davies normalization). To first order in slow roll, we have:
\bea
u(\tau,k)= \frac{i\, H^2}{\sqrt{4 \gamma_1 c_s^3 k^3}}(1+i k c_s \tau)e^{-i k c_s \tau},\qquad  {\rm with} \quad \zeta= -H \xi = \int \frac{d^3 p}{(2\pi)^3}\, u(\tau,p)\, e^{i \vec p \cdot \vec x } .
\eea

\noindent Without much ado we proceed to the calculation of the various contributions to $f_{NL}$. 
This has been done for the single-field case in \cite{Andrew1} and, considering our use of the $SO(N)$ symmetric to obtain an effectively single degree of freedom behaviour, we simply and briefly produce here  a discussion of the shape-function, and plot it.  The formalism used is the Schwinger-Keyldish\cite{ININ1,ININ2,ININ3}, so called IN-IN, formalism (see the Appendix for further details).

\noindent The two-point function for the observable $\zeta$ is defined as:
\bea
\langle \zeta(\tau,k_1) \zeta(\tau,k_2) \rangle= (2\pi)^3 \delta^{(3)}\left(\vec k_1+\vec k_2\right) P({k_1}), 
\eea
we also introduce $P_{\zeta}=P_{k} \, k^3/(2\pi^2) $, whose value is found from COBE normalization \cite{Smoot:1992td}. In the two-point function calculations the interactions enter only at the loop level while here we focus on tree level diagrams. One therefore moves to the three-point function, where cubic interactions enter at tree level. The part of the action cubic in the fluctuations is:
\bea
S_3(\xi)=\int d^4 x\, \alpha_4\,a^3 Z \frac{\dot{\bar{\pi}}^3 }{\Lambda^3} \Bigg[18 H   \dot{\xi}^3-12    {\dot \xi}^2 \p_i^2 \xi  -14 H  {\dot \xi} (\p_i \xi)^2+3 (\p_i \xi)^2  \p_j^2 \xi     \Bigg] \,\,. 
\eea

We are after the three-point correlator shape-function, we find it by isolating a momentum conservation Dirac delta from the three-point function:
\bea
\langle \zeta(\tau,k_1) \zeta(\tau,k_2) \zeta(\tau,k_3)  \rangle= (2\pi)^3\delta^{(3)}(\vec k_1+\vec k_2+\vec k_3) B(\tau, k_1,k_2,k_3)
\eea
The amplitude $f_{NL}$ is defined as:
\bea
f_{NL}\equiv\frac{5}{6} \frac{B(\tau, k_1,k_2,k_3)}{\big[P(k_1) P(k_2)+\, perm.\big]} \quad,
\eea
and then, using invariance under the overall size of the $|\vec k|$, one plots the function
\bea
 k_2^2\, k_3^2\times  B\left(\frac{k_1}{k_1} , \frac{k_2}{k_1}, \frac{k_3}{k_1}\right),
\eea
\noindent where the first factor has been added to enhance the differences in the shape-functions profile.

\begin{figure}[htbp]
\begin{center}
\includegraphics[scale=0.70]{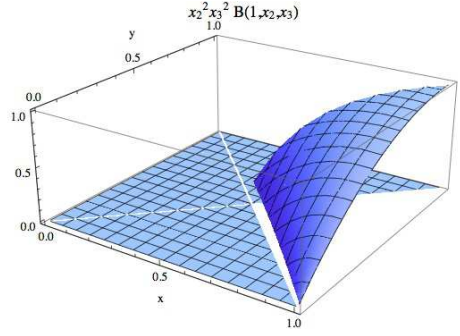}
\caption{The shapefunction peaks in the equilateral ($k_n/k_1\equiv x_n=1$) configuration, as expected.}
\label{default1}
\end{center}
\end{figure}

\noindent Whenever a theory has many independ coefficients regulating the interactions one obtains different sorts of shapefunctions by tuning the different coefficients accordingly. Not so in this case, the $SO(N)$ symmetry, no matter the value of $c_s$ (which depends also on $\alpha_2$, not just $\alpha_4$), demands an equilateral shapefunction, we plot it in Fig~(\ref{default1}).

As for the value of $f_{NL}$ in this setup, it will depend on both $\alpha_4$ and $\alpha_2$ or, which is the same, $\alpha_4$ and $c_s$. But, considering that the value of $c_s$ in most of the parameters space of this model resides\footnote{Unless one is willing to fine-tune $\alpha_{2,4}$ to obtain very small values for $c_s$, one can only assume\\ $\alpha_2\sim \alpha_4\, ,or \,\,\,\alpha_2\gg\alpha_4\, , or\,\,\, \alpha_2\ll\alpha_4\, .$ } in the interval $\big[1/2,1 \big]$, one could quickly conclude that value of $|f_{NL}|$ is less than unity. In fact, a detailed calculation shows (see also \cite{Andrew1}) that in the so-called equilateral ($k_1=k_2=k_3\equiv k$) configuration the bispectrum amplitude is:
\bea
f_{NL}= \frac{20 \alpha_{4} \left(17 \alpha_{2}^2+835 \alpha_{2} \alpha_{4}+2574 \alpha_{4}^2\right)}{81 (\alpha_{2}+26 \alpha_{4})^2 (\alpha_{2}+54 \alpha_{4})}+ {\rm s.r.}\,\,,
\eea
\noindent where we have given the expression at leading order in slow roll and we have set $Z=1$. \footnote{Remember, $Z\sim 1$ because we want the Galilean non-linearities to be important whilst at the same time not running into strong coupling issues \cite{Andrew1}.} 

A straighforward study of the $\alpha_2,\alpha_4$-dependence of $|f_{NL}|$ reveals that, with the exception of a very small region in the $\alpha_2,\alpha_4$ plane \footnote{This  is an extremely small region in the $\alpha_{2,4}$ plane, furthermore one can show it corresponds to a very small $c_s$ and it can therefore be disregarded on the sole basis that, as well known, a very small $c_s$ takes us into realms where perturbation theory might break down.}, the quantity above is always smaller than unity.

\noindent We now proceed to the trispectrum calculation. The quartic contribution to the fluctuations Hamiltonian\footnote{We aim directly at the Hamiltonian in this case because, as opposed to the cubic case, $\mathcal{H}_4 \not= -\mathcal{L}_4$ and therefore one needs to be more careful \cite{Huang:2006eha}.} is now needed:
\bea
\mathcal{H}_4=\,\frac{a^3}{\dot{\bar{\pi}}^4}  \Bigg[ \mathcal{O}_{1}   \delta \dot{\pi}^4   + \mathcal{O}_{2}   \delta \dot{\pi}^3 \frac{\p_i^2 \delta\pi}{a^2} + \mathcal{O}_{3}\delta \dot{\pi}^2 \frac{(\p_i \delta\pi)^2}{a^2}  + \mathcal{O}_{4}\delta \dot{\pi}^2 \frac{(\p_i^2 \delta\pi)^2}{a^4}  +  \mathcal{O}_{5}\delta \dot{\pi}^2 \frac{(\p_i \p_j \delta\pi)^2}{a^4}
\nonumber \\
+ \mathcal{O}_{6} \delta \dot{\pi} \frac{(\p_i \delta \pi)^2}{a^2} \frac{\p_j^2 \delta \pi}{a^2}  +\mathcal{O}_{7} \frac{(\p_i \delta\pi)^4}{a^4} +  \mathcal{O}_{8}\frac{(\p_i \delta\pi)^2}{a^2}\cdot \frac{\left( (\p_j^2 \delta\pi)^2-\p_k\p_l \delta \pi \p_k\p_l \delta \pi    \right)}{a^4} \Bigg] \nonumber \\ \label{H}
\eea 
where, to first order in slow roll, $\zeta,\xi,\delta \pi$ are simply related by $\zeta=-H \xi= -(H/ \dot{\bar{\pi}}) \delta \pi $. The expression for $\mathcal{H}_4$ with the explicit $\mathcal{O}_n$ operators it provided in the \textit{Appendix}.

\noindent The general formula for the calculation of higher order correlators in the IN-IN formalism is given by:
\bea
 \langle \Omega|\zeta_{k_1}\zeta_{k_2}\zeta_{k_3}...\zeta_{k_n}(t)|\Omega \rangle = \langle 0|\bar T \{e^{i\int_{-\infty}^{t} d^3 x dt^{'} \mathcal{H}_I(x)}\} \zeta_{k_1}\zeta_{k_2}\zeta_{k_3}...\zeta_{k_n}(t)\,  T\{ e^{-i\int_{-\infty}^{t} d^3 x^{'} dt^{''} \mathcal{H}_I(x)} \}|0\rangle \nonumber ,\\  \label{ttb}
\eea
\noindent we spare the reader the details of the analytical results of the calculation in the main text, focusing here on the plots of the shape function (also known as form factor); we report them in the \textit{Appendix}. The trispectrum in Fourier space is also defined as:
\bea
\langle \zeta({\tau,k_1}) \zeta({\tau,k_2})\zeta({\tau,k_3})\zeta({\tau,k_4}) \rangle = (2\pi)^9 P_{\zeta}^3\, \delta^{(3)}\left(\sum_{i=1}^{4} k_i \right) \prod_{i=1}^{4} \frac{1}{k_i^3}\mathcal{T}(k_1,k_2,k_3,k_4,k_{12},k_{14}) \nonumber \\
\eea
\noindent where $\mathcal{T}$ is the form factor we will plot. Notice that the four-point correlator shape-function depends on six parameters so one has to choose different $k$-configurations for a plot. We choose them here in order to ease the comparison with existent literature with specific attention to the results of other stable (in the sense of effective field theory) inflationary models, such as DBI inflation. Our main reference for comparison will be \cite{Chen:2009bc}.

\begin{figure}[htbp]
\begin{center}
\includegraphics[scale=0.39]{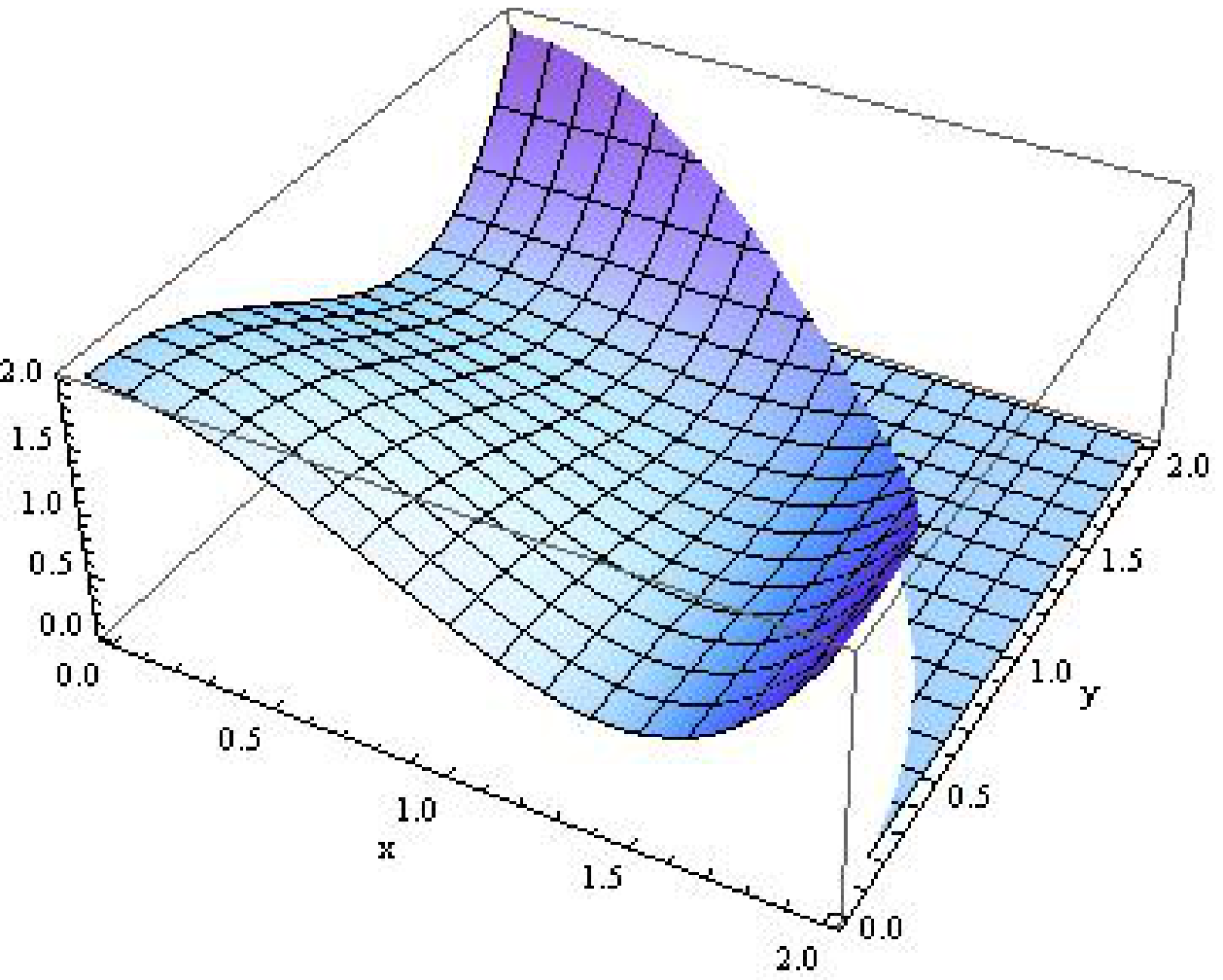}
\hspace{3cm}
\includegraphics[scale=0.34]{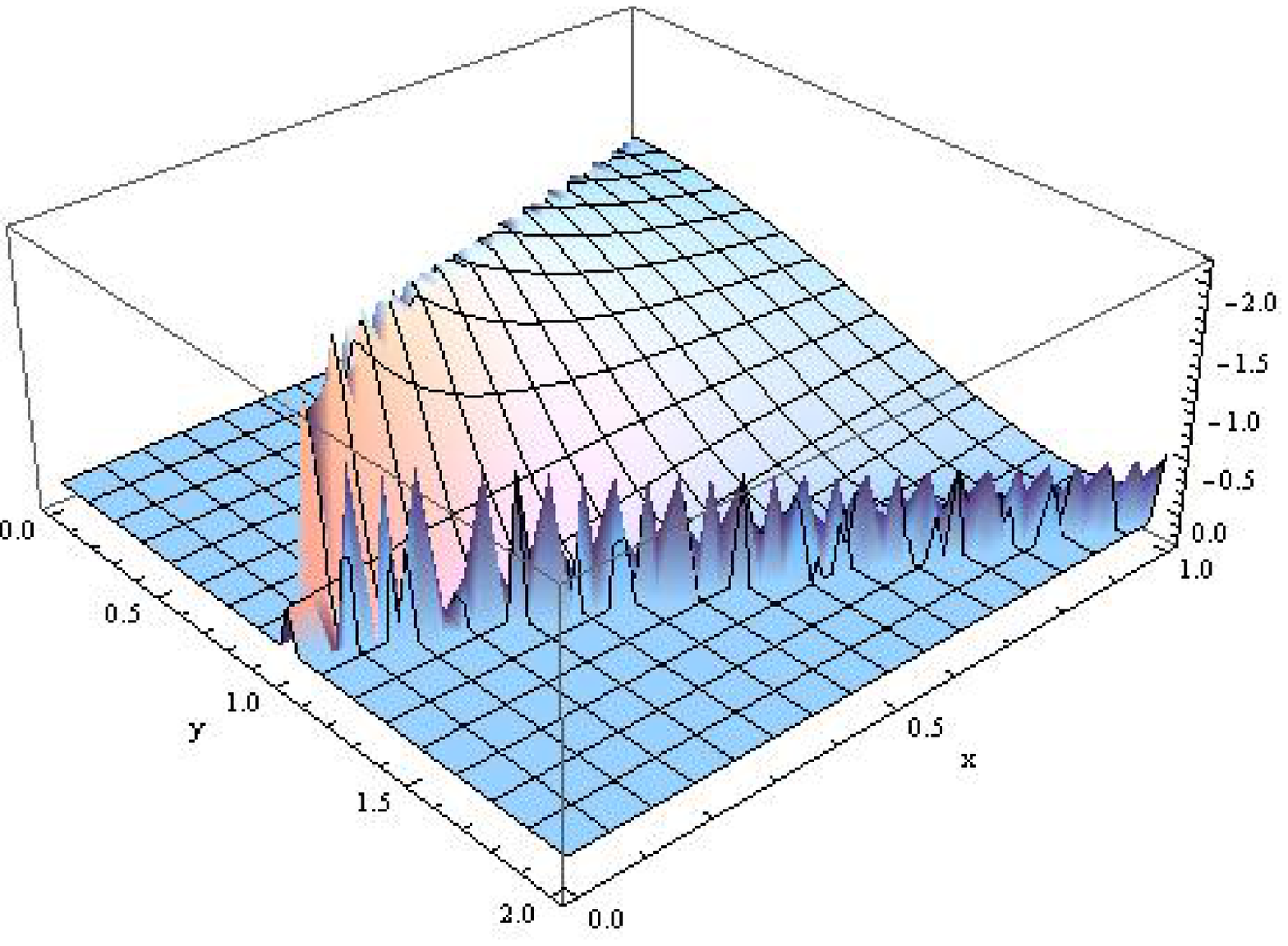}
\caption{The equilateral configuration on the left strongly differs from the DBI result in that, for example the $(0,0)$ value is finite, non-zero and it is actually the peak of the shape function. This is to be compared with a  $\mathcal{T}$ which has its minimum at $(0,0)$ in $P(X,\pi)$ theories. \newline
Also, the double squeezed configuration plot on the right shows  a finite non-zero shape function in the $(k_{12} \rightarrow 0,\,\,\, y=k_{14}=1)$ limit. This is only to be found in contributions to the trispectrum coming from third order interaction terms in $P(X,\pi)$ models.  On the contrary, what is plotted here is a so called \textit{contact interaction} contribution, thus coming from the fourth order perturbations.}
\label{default3}
\end{center}
\end{figure}

\noindent Pictorially, the k-configurations can be understood by looking at the tetrahedron below:

\begin{figure}[htbp]
\begin{center}
\includegraphics[scale=0.60]{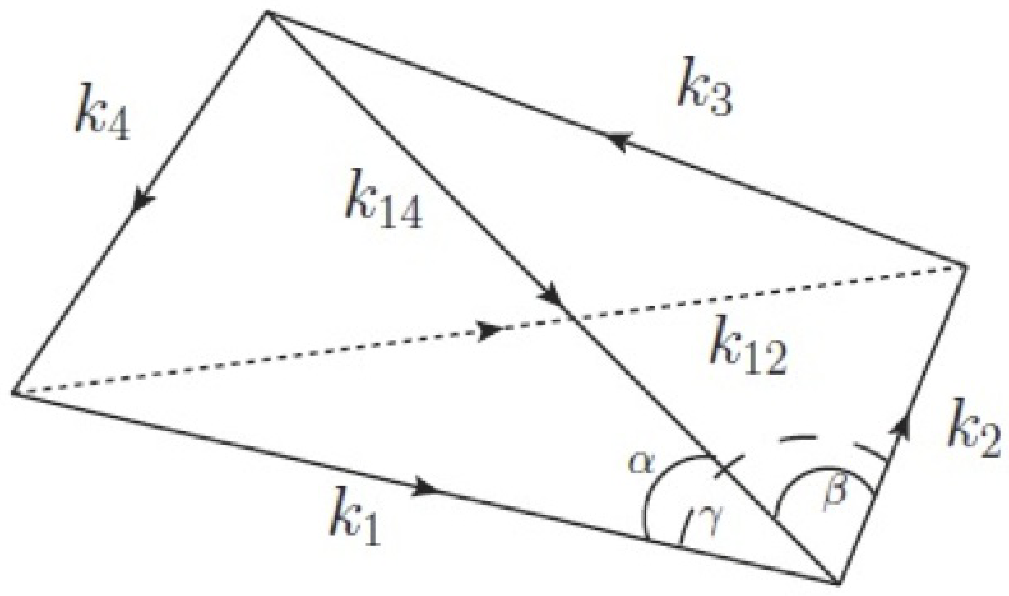}
\caption{\newline Equilateral configuration: $k_1=k_2=k_3=k_4$, plotting  $k_{12}\equiv |\vec k_1 +\vec k_2|$, $k_{14}\equiv |\vec k_1 +\vec k_4|$. $\qquad \qquad\qquad\qquad$
 Folded configuration : $k_{12} \rightarrow 0, \quad k_1=k_2, \quad k_3=k_4$, plotting $k_{14}/k_1,\,\,\, k_4/k_1$ .
 \newline Specialized planar limit: $k_1=k_3=k_{14}$; $k_{12}=f(k_1,k_2,k_4)$ , plotting $k_2/k_1$ , $k_4/k_1$. \newline Near double squeezed limit: $k_3=k_4=k_{12}$; $k_2=g(k_1,k_3,k_4,k_{14},k_{12})$  , plotting $k_{14},k_4$}
\label{default2}
\end{center}
\end{figure}
\noindent What one plots is the $\mathcal{T}$ form factor, except in the last configuration, where $\mathcal{T}\cdot \prod_{i} 1/k_i$ is preferred, for an enhanced effect. We report below on the plots we obtain for our model. Interestingly, both in the equilateral and the double squeezed configurations the shape-function is distinctively different from its $P(X,\pi)$ counterpart. The most striking differences  are detailed in the caption of Fig.~(\ref{default3}).

\begin{figure}[htbp]
\begin{center}
\includegraphics[scale=0.39]{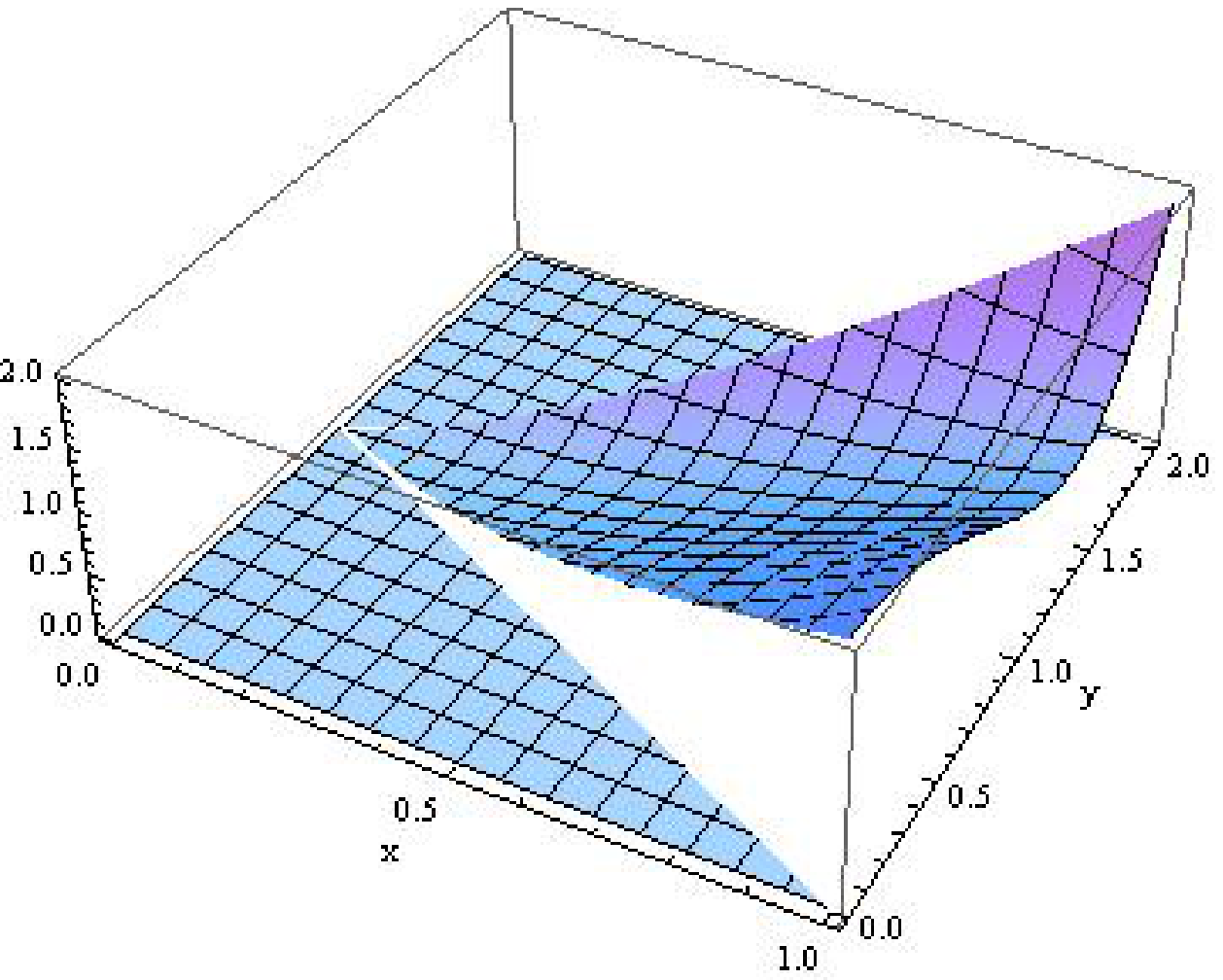}
\hspace{3cm}
\includegraphics[scale=0.39]{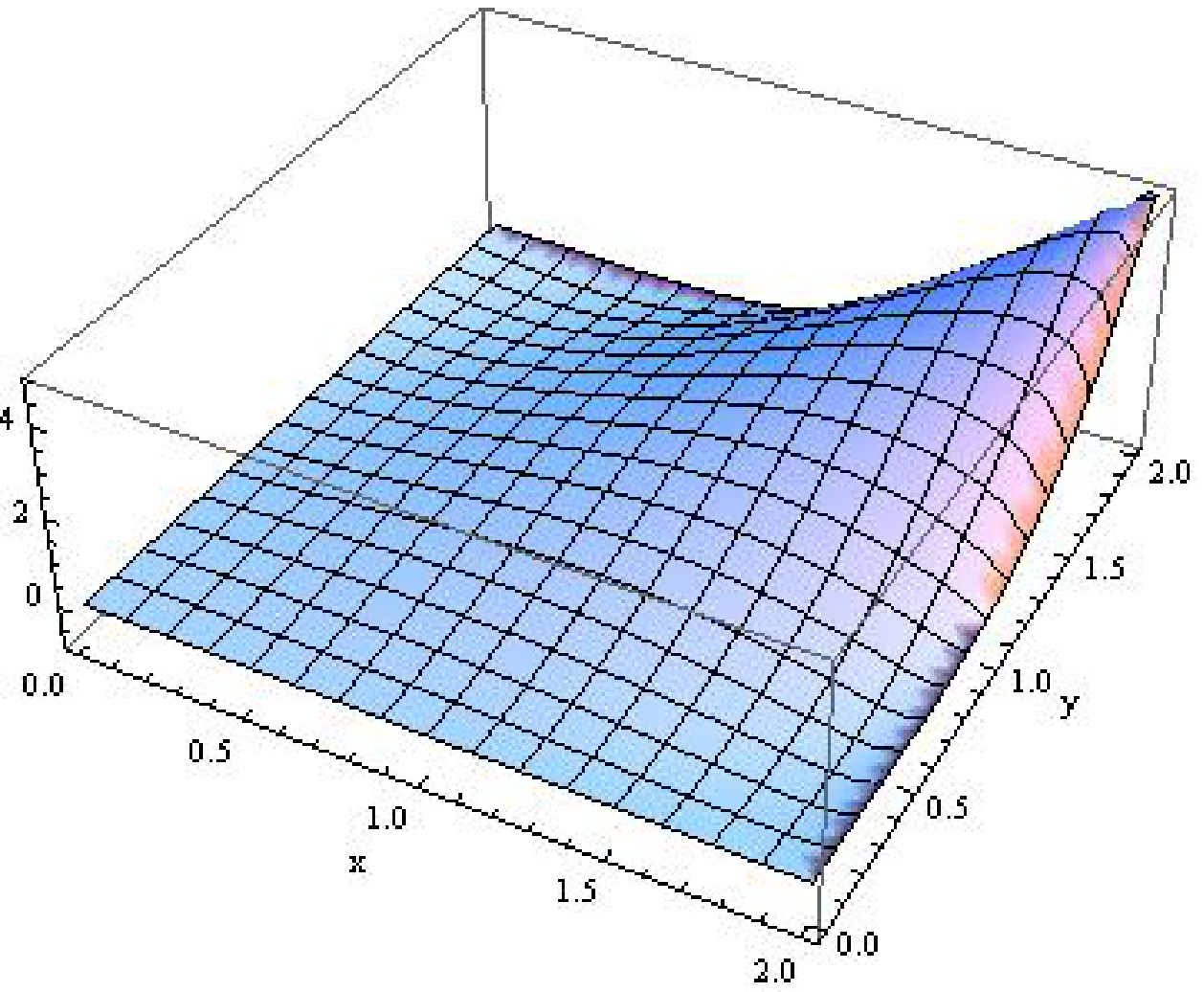}
\caption{The folded configuration form factor is plotted on the left. The planar limit shape is on the right side. These configuration do not have particularly distinctive signatures as compared to the corresponding plots in \cite{Huang:2006eha}}
\label{default4}
\end{center}
\end{figure}

 As of today, we do not have as precise a bound on any given trispectrum configuration as we do on the bispectrum side ($f_{_{NL}} \in \{f^{loc}_{NL},f^{eq}_{NL},f^{ortho}_{NL} \}$) but, if we were to reach such a precision, these shape-functions differences could be translated into something immediately testable. 
 As for the other two configuration, they do not carry as much information as the first two in that there the shapes are hardly distinguishable from the $P(X,\pi)$ ones. We report them in Fig.~(\ref{default4}) for completeness.

\noindent All the shapes plotted in Fig.(\ref{default3},\ref{default4}) are obtained for the  $c_s\sim 1$ case, in each configuration.  One way to obtain $c_s\sim 1$ is to assume $\alpha_2\gg \alpha_4$, an inequality\footnote{It is important to note that in no way this condition is in conflict with our working in the regime where non-linearities are important. In fact, our inequality would be better written as $\alpha_2\cdot  \alpha_4\gg\alpha_4^2$ , that is, $\alpha_4$ is still playing an important role.} which also automatically grants that the \textit{scalar exchange}\footnote{In the Feynman diagrams language, the contact interaction contribution is the single-vertex interaction coming from $\mathcal{H}_4$, the scalar exchange one originates from two vertices in $\mathcal{H}_3$, both are tree level interactions. } contribution to the trispectrum will be subdominant. 

The $c_s\sim 1/\sqrt{2}$ (as low as one can go in $c_s$ without strong fine tuning) plots on the other hand, do involve the calculation of the scalar exchange contribution. This is the $\alpha_4\gg \alpha_2$ regime.
At this stage one would usually plot for this regime both the \textit{scalar exchange} contribution and the \textit{contact interaction} one separately.

\noindent If we were to proceed according to this prescription,
we would find that both \textit{c.i.} and \textit{s.e} contributions to the shape-function distinctively differ from $P(X,\phi)$ models in the equilateral configuration. The \textit{c.i.} also has novel features in the double squeezed configuration, much like the case of Fig.~(\ref{default3}, right).

In this theory though, we have just one coupling and, $\alpha_2$ being subleading in this region of the parameters space, we can simply sum all the contributions to the overall trispectrum. The resulting shape-functions are plotted below in Fig.~(\ref{default5}). 

As before,  folded and planar configurations are not at all illuminating, the trispectrum profile there essentially overlaps with the results plotted above and with those of  \cite{Huang:2006eha}. We therefore do not report them in the text. The double squeezed limit plot requires some clarification: since we are summing the \textit{c.i.} and \textit{s.e} contributions, we should not be looking for ways to distinguish their respective profile as is generally done elsewhere, in less restricted (which is tantamount to -symmetric- here) models.

The equilateral configuration, despite the fact that here too we are summing the various \textit{c.i.} and \textit{s.e} contributions, shows a profile which one can easily see is hardly obtainable by summing \textit{c.i.} and \textit{s.e} terms in $P(X,\phi)$-type theories.

As for the trispectrum amplitude $\tau_{NL}$, it is defined in direct analogy to $f_{NL}$, as:\bea
\large{\langle \zeta^4 \rangle_{reg.}
\rightarrow_{} (2\pi)^9 P_\zeta^3 \delta^3 (\sum_i
\vec k_i) \frac{1}{k^9} \, \tau_{NL} \,, }
\eea
to be calculated in the regular tetrahedron limit. 
Its specific analytical expression  for the model studied here is 
\bea
\tau_{NL}=\frac{\alpha_{4} \left(1.2\, \alpha_{2}^5+2.8 \times10^2\, \alpha_{2}^4 \alpha_{4}+2.2\times 10^4\, \alpha_{2}^3 \alpha_{4}^2+7.3\times10^5 \alpha_{2}^2 \alpha_{4}^3+1\times 10^7 \alpha_{2} \alpha_{4}^4+6.7\times 10^7 \alpha_{4}^5\right)}{(\alpha_{2}+26\, \alpha_{4})^4 (\alpha_{2}+54\, \alpha_{4})^2}. \nonumber \\
\eea

 A straighfotrward study of its dependence on the parameters of the theory reveals that it too is smaller than unity. Again, there exist a very small region in the $\alpha_2,\alpha_4$ plane  where $\tau_{NL}$ becomes larger, but it corresponds to a region where $c_s$ is extremely small. 


\begin{figure}[htb]
\begin{center}
\includegraphics[scale=0.46]{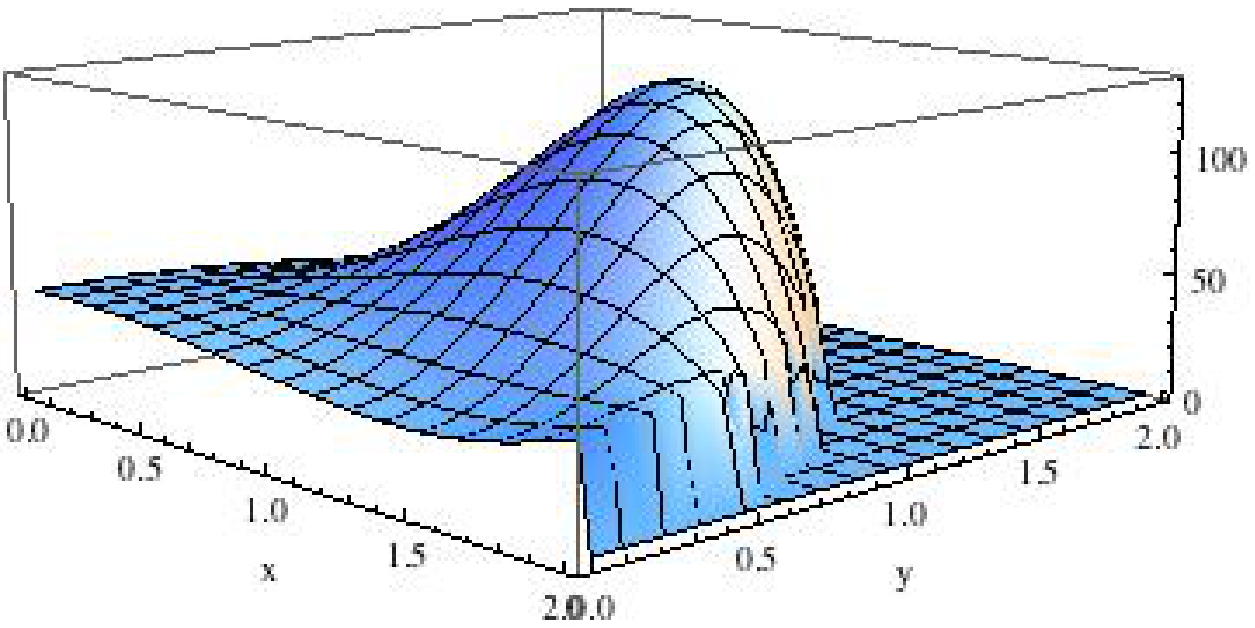}
\hspace{3cm}
\includegraphics[scale=0.42]{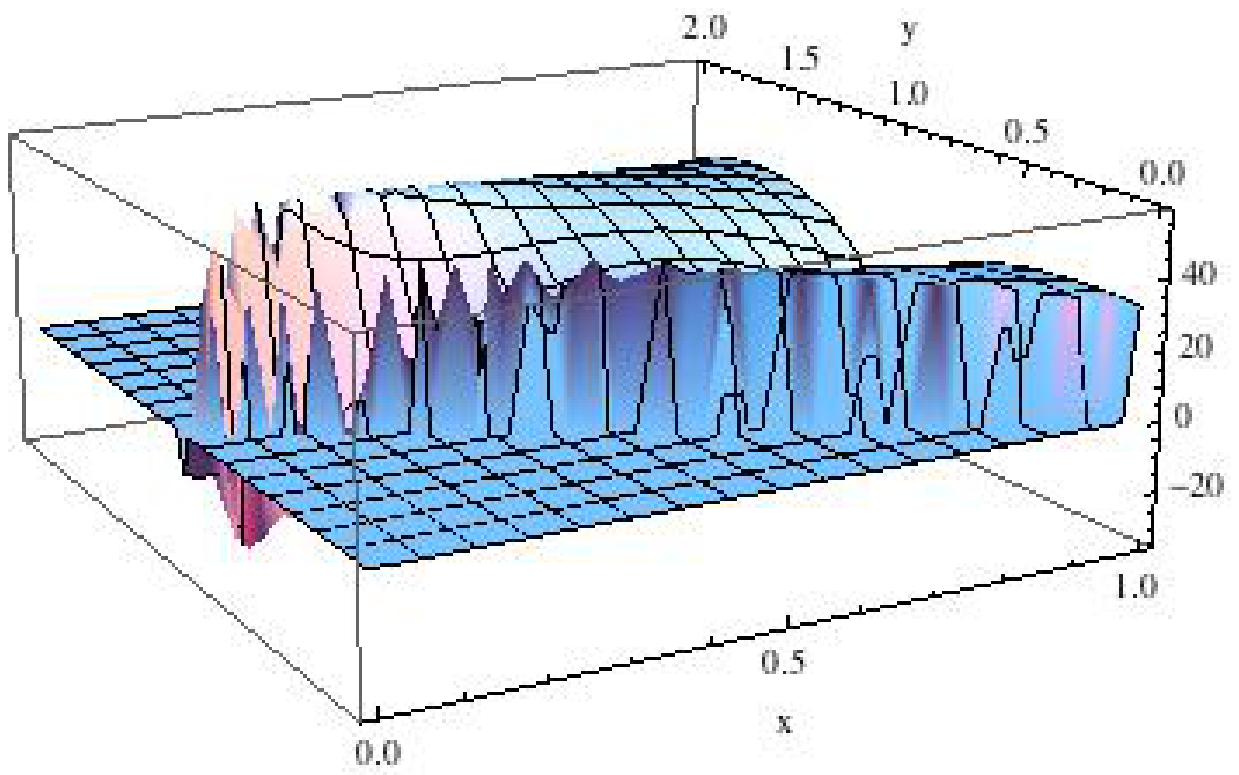}
\caption{The equilateral configuration on the left: although a comparison to the $P(X,\pi)$ theories counterpart is less straightforward, one can still observe that the profile here is such that it would not be possible to reproduce it in, say DBI inflation. \newline
The double squeezed configuration on the right: see main text for further comments}
\label{default5}
\end{center}
\end{figure}

What we have here then is a quite distinct model of Galileon inflation which shares with the related co-dimension 1 model \cite{Andrew1} all the essential stability features. The additional $SO(N)$ symmetry has implemented a theory with only $\alpha_2, \alpha_4$ as parameters and resulted in a bit less flexibility when it comes to small $c_s$ and correspondingly large $f_{NL}$. Nevertheless, the model accommodates for a $f_{NL} \lesssim 1$ and  has quite intriguing trispectrum signatures. We further expand on this point in the conclusions. 

We now offer some comments on the small $f_{NL}$ vs large $\tau_{NL}$  possibility; this is a topic which, when looked at  from a quantum field theory perspective, is interesting beyond the cosmological perturbations framework.

\section{Discussion}
As we have see above, this symmetry dictated Galileon model is quite interesting on its own as it produces signatures in the trispectrum shape function which clearly distinguish it from the $P(X,\pi)$ (chief among which, DBI inflation) predictions.

Precisely because of this symmetry, the full cubic Lagrangian is absent from the onset. If such an occurrence were to even mildly (one only needs or maybe prefers a parametrically small three-point function, rather than a null one) propagate itself into the action for perturbations, it would be quite intriguing. This is  so because the consequently small $f_{NL}$ parameter would imply a $\tau_{NL}$ at  close observational reach. If Planck data will not rule out a $f_{NL}$ compatible with zero or if a small $f_{NL}$ is the most favored one, the very next natural step will be to deepen the study of trispectrum shape-functions and tighter bounds on $\tau_{NL}$ would be in sight \footnote{Strictly speaking this is, of course, a model-dependent statement.}.

How then does one obtain in a natural way the combination of a small $f_{NL}$ with a large $\tau_{NL}$? There are some ideas in the literature, most notably \cite{Senatore:2010jy,Izumi:2010wm} (see \cite{Bartolo:2010di} for an implementation of these in the study of the trispectrum of very generic effective theories), which all approach inflation within the effective field theory of inflation methods of \cite{Cheung}. 

Assuming that there exists a clear-cut symmetry such as e.g. $\delta \pi \rightarrow -\delta \pi$ of \cite{Senatore:2010jy}, immediately serves our purpose, but plenty of work is still ahead if one wants to arrive at the full effective theory. Indeed, it is necessary to reverse engineer the fluctuations action; the process is not easy and certainly not unique. What appears to be a straightforward symmetry in the  theory of fluctuations, might reveal a very contrived expression in the full one \cite{pert_full}.\\

\noindent This is  a quite generic issue in inflationary setups: \\

\noindent- there are models which are openly phenomenologically inspired and so are not thought for nor suitable  to analyze issues such like unitarity and stability of the theory.\\

\noindent- there are theories which are unitary and stable (e.g. \cite{DBI, Andrew1}) and therefore are known in their full, background independent, form. Their predictions are about to be put to more stringent tests by Planck data.\\

\noindent - there is an effective theory of fluctuations approach \cite{Cheung}, which is extremely powerful in reading off what happens at the fluctuations level and mapping it onto specific operators in the Lagrangian $\mathcal{L}(\delta\pi^n)$. In this approach one rightfully assumes that there is an FRW background to expand around and cleverly uses it to gain calculational advantage.  The clear-cut dictionary between non-Gaussianities and specific independent operators is ideally suited for the task of predicting almost all possible non-Gaussian scenarios that data might show. 

However, it is quite hard to further map the $\mathcal{L}(\delta\pi^n)$ symmetries into their full $\mathcal{L}(\pi^n)$ counterpart. \\

\noindent - an effective background independent field theory of inflation has also been put forward \cite{Weinberg:2008hq}. The Galileon model discussed here nicely fits within this setup (DBI inflation goes even further in that it can claim UV finite status). It is in this approach that the properties of the full theory are more manifest and, also, where the theory is in a form more suitable to guess any corresponding UV finite embedding.\\

\noindent But, as we have seen, it is hard to implement in a simple way seemingly straightforward requirements from the non-Gaussianities endpoint of the theory. Further inputs might materialize with the upcoming release of Planck data and it might soon be of utmost importance to create a dictionary between this approach and the one of \cite{Cheung} if we are after a stable model of inflation which is not accommodated by any of the models in e.g. \cite{DBI, Andrew1}. 

Even if  in cosmological setups $f_{NL}$ in all its probed disguises (local, equilateral, orthogonal etc..) turns out to be accounted for by existing models, this 
\[
\mathcal{L}(\delta\pi^n)\,\,  \underleftrightarrow{symmetries}\,\, \mathcal{L}(\pi^n) \nonumber
\]
 dictionary  is an interesting quantum field theory issue by its own \cite{pert_full}. Everytime one has, as is frequently the case, a window on perturbations around a specific background and a symmetry is apparent in that setup, it is not at all obvious how to retrace that into the full theory (when available). We plan to expand on this and provide concrete examples of a dictionary in \cite{pert_full}.

\section{Conclusions}
We have studied a specific inflationary model in a regime where non-linear Galileon interactions play a leading role.   This theory stems from the study of the higher dimensional brane constructions origins of Galileon interactions, much in the spirit of the original DGP proposal \cite{DGP}. Whenever the dimensionality of the (bulk -brane) space is even, if the Galileon interactions are to preserve their essential flat space and co-dimension $1$ properties, an additional SO(N) symmetry arises which greatly restricts the allowed building blocks in the theory. Indeed, in the $4D$ (3-brane) case, only one coupling regulates the interactions.

Having a smaller parameters space translates into less flexibility when it comes to the value of the speed of sound $c_s$ and non-Gaussianities in general. However, we have shown that the bispectrum amplitude can go as high as  $f_{NL}\lesssim 1$ and peaks in the equilateral configuration. We have used the $SO(N)$ symmetry to consistently limit the dynamics to a single-field case by choosing appropriate initial conditions.

The trispectrum shape-function is quite intriguing in that it is distinctively different than the corresponding $P(X,\pi)$ results in more than one momenta configuration: equilateral and double squeezed.

We stress here that the galilean origin of this theory endows it with second order equations of motion (unitarity at the quantum level) and non-renormalized interactions. The latter makes sure the number of leading coefficients in a specific regime stays the same, an issue which is typically highly non trivial in generic inflationary models, and makes the theory under study properly predictive.  These are the grounds on which we have chosen to compare the predictions of the model against $P(X,\pi)$ theories,  with especially $DBI$-inflation in mind.

Despite in this model all odd (in the scalar field $\pi$) Lagrangian contributions are null in the full theory, we have shown that there is no room for a small bispectrum \textit{vs} large trispectrum combination to occur. This fact has prompted some comments on how to realize such a scenario and what it means to translate symmetries in the fluctuations Lagrangian into their full theory counterparts. This will be essential if Planck data suggests such a possibility and represents an interesting question on its own.

Besides upcoming enhanced sensitivity on cosmological observables, a way to navigate the ever-growing space of inflationary models compatible with data is to be more demanding on the effective quantum field theory side, taking an inflationary model seriously enough to request quantum stability, if not UV finite origins. Galileon inflation sits nicely in this context and we hope to further investigate related models in the future.\\

\noindent {\bf Note added}: this work has been performed before the release of Planck mission data. In light of the results in \cite{Ade:2013ydc} ($f_{NL}$ small in all the probed spectrum) the results obtained here (a small $f_{NL}$ indeed) still stand. It would also be of particular interest now the possibility of a naturally small $f_{NL}$ combined with a large $\tau_{NL}$, a topic on which we have offered some comments here and hope to offer some results in the future.

\acknowledgments

It is a pleasure to thank A.~J.~Tolley for many enlightening discussions and F.~Arroja, N. ~Bartolo, E.~Dimastrogiovanni for fruitful and stimulating conversations. MF is supported in part by D.O.E. grant DE-SC0010600.

\section{Appendix}
We report below the analytical result of the different contributions to the trispectrum generated by the interactions in Eq.~(\ref{H}). \\
We organize them by operator and therefore have seven different contributions (two of the eight different interactions in Eq.~(\ref{H}) have identical analytical expression.

\bea
T_{\zeta}^{(1)}&=&- \left(\frac{H^{12}\mathcal{O}_{1}}{\gamma_{1}^{4}c_{s}^{9}}\right)\left(\frac{3}{16\,k_{t}^{5}\, \Pi_{i}k_{i}}\right)=\frac{9}{4}\frac{H}{c_{s}^{2}}\frac{\mathcal{O}_{2}}{\mathcal{O}_{1}}T_{\zeta}^{(2)}\nonumber ,\\ \nonumber
T_{\zeta}^{(3)}&=&- \left(\frac{H^{12}\mathcal{O}_{3}}{\gamma_{1}^{4}c_{s}^{11}}\right)\left(\frac{\hat{k}_{3}\cdot\hat{k}_{4}}{16\,k_{t}^{5}\, \Pi_{i}k_{i}}\right)\left(3+\frac{3k_{t}}{4k_{3}}+\frac{3k_{t}}{4k_{4}}+\frac{k_{t}^{2}}{4k_{3}k_{4}}\right)+\,23\,\,perms.\,,\\
T_{\zeta}^{(4)}&=& -\left(\frac{H^{14}\mathcal{O}_{4}}{\gamma_{1}^{4}c_{s}^{13}}\right)\left(\frac{3}{16\,k_{t}^{5}\, \Pi_{i}k_{i}}\right)\left(1+\frac{5k_{3}}{k_{t}}+\frac{5k_{4}}{k_{t}}+\frac{30k_{3}k_{4}}{k_{t}^{2}}\right)+\,23\,\,perms.\, , \nonumber \\ \nonumber
T_{\zeta}^{(5)}&=&- \left(\frac{H^{14}\mathcal{O}_{5}}{\gamma_{1}^{4}c_{s}^{13}}\right)\left(\frac{3\left(\hat{k}_{3}\cdot\hat{k}_{4}\right)^{2}}{16\,k_{t}^{5}\, \Pi_{i}k_{i}}\right)\left(1+\frac{5k_{3}}{k_{t}}+\frac{5k_{4}}{k_{t}}+\frac{30k_{3}k_{4}}{k_{t}^{2}}\right)+\,23\,\,perms.\,  ,\\
T_{\zeta}^{(6)}&=&- \left(\frac{H^{13}\mathcal{O}_{6}}{\gamma_{1}^{4}c_{s}^{13}}\right)\left(\frac{\hat{k}_{3}\cdot\hat{k}_{4}}{16\,k_{t}^{5}\, \Pi_{i}k_{i}}\right)\(3+\frac{3k_{2}}{k_{3}}+\frac{3k_{2}}{k_{4}}+\frac{15k_{2}}{k_{t}}+\frac{3k_{t}}{4k_{3}}+\frac{3k_{t}}{4k_{4}}+\frac{3k_{2}k_{t}}{4k_{3}k_{4}}+\frac{k_{t}^{2}}{4k_{3}k_{4}}\)+\,23\,\,perms.\, \nonumber ,\\ \nonumber
T_{\zeta}^{(7)}&=& -\left(\frac{H^{12}\mathcal{O}_{7}}{\gamma_{1}^{4}c_{s}^{13}}\right)\left(\frac{(\hat{k}_{1}\cdot\hat{k}_{2})(\hat{k}_{3}\cdot\hat{k}_{4})}{64\,k_{t}^{4}\, \Pi_{i}k_{i}^{2}}\right)\left(k_{t}^{3}+3\sum_{i<j<k}k_{i}k_{j}k_{k}+3k_{t}\sum_{i<j}k_{i}k_{j}\right)+ \,23\,\,perms.\,,\\
T_{\zeta}^{(8)}&=&- \left(\frac{H^{14}\mathcal{O}_{8}}{\gamma_{1}^{4}c_{s}^{15}}\right)\frac{\hat{k}_{1}\cdot\hat{k}_{2}\left(1-(\hat{k}_{3}\cdot\hat{k}_{4})^{2}\right)}{16\,k_{t}^{5}\, \Pi_{i}k_{i}}\left(\frac{k_{t}^{2}}{k_{1}k_{2}}+3\,\sum_{i<j}\frac{k_{i}k_{j}}{k_{1}k_{2}}+15\,\sum_{i<j<k}\frac{k_{i}k_{j}k_{k}}{{k_{1}k_{2}}k_{t}^{}}+\frac{90k_{3}k_{4}}{k_{t}^{2}}\right) +\,23\,\,perms.\,,\nonumber \\
\eea
where $k_{t}\equiv k_{1}+k_{2}+k_{3}+k_{4}$ and

\bea
\mathcal{O}_{1}\equiv-\frac{9\,\alpha_{4}H^{2}(\alpha_{2}-270\,\alpha_{4}Z^{2})\dot{\bar{\pi}}^{6}}{4\,\gamma_{1}\Lambda^{6}} ; \quad \mathcal{O}_{2}\equiv \frac{4\,\alpha_{4}H(\alpha_{2}-270\,\alpha_{4}Z^{2})\dot{\bar{\pi}}^{6}}{2\,\gamma_{1}\Lambda^{6}};\\
\mathcal{O}_{3}\equiv \frac{7\,\alpha_{4}H^{2}(\alpha_{2}-54\,\alpha_{4}Z^{2})\dot{\bar{\pi}}^{6}}{2\,\gamma_{1}\Lambda^{6}}; \quad 
\mathcal{O}_{4}\equiv -\frac{3\,\alpha_{4}(\alpha_{2}-42\,\alpha_{4}Z^{2})\dot{\bar{\pi}}^{6}}{2\,\gamma_{1}\Lambda^{6}} ;
\eea
\bea
\mathcal{O}_{5}\equiv \frac{3\,\alpha_{4}\dot{\bar{\pi}}^{4}}{\Lambda^{6}}; \quad
\mathcal{O}_{6}\equiv -\frac{6\,\alpha_{4}H(\alpha_{2}-2\,\alpha_{4}Z^{2})\dot{\bar{\pi}}^{6}}{2\,\gamma_{1}\Lambda^{6}} ;\\
\mathcal{O}_{7}\equiv -\frac{2\,\alpha_{4}H^{2}(\alpha_{2}+5\,\alpha_{4}Z^{2})\dot{\bar{\pi}}^{6}}{2\,\gamma_{1}\Lambda^{6}}; \quad 
\mathcal{O}_{8}\equiv\frac{\alpha_{4}\dot{\bar{\pi}}^{4}}{\Lambda^{6}}.
\eea

\bea
&\mathcal{H}_4=\frac{\alpha_{4} a^3 \dot{\bar{\pi}}^2}{4\gamma_{1}\Lambda^{6}}\Bigg[\alpha_{2}\left(-9 D_1 H^2 + \frac{8 D_2 H}{c_s^2} + \frac{14 D_3 H^2}{c_s^2} - \frac{6 D_4}{c_s^2} + \frac{
 6 D_5}{c_s^4} - \frac{12 D_6 H}{c_s^4} - \frac{4 D_7 H^2}{c_s^4} + \frac{2 D_8}{c_s^6}\right)\qquad \qquad\qquad \qquad\qquad \quad \qquad \nonumber \\
 &+\alpha_{4}\left(2430 D_1 H^2 Z^2-\frac{2160 D_2 H Z^2}{c_s^2} -\frac{
 756 D_3 H^2 Z^2}{c_s^2}+\frac{252 D_4 Z^2}{c_s^2} + \frac{324 D_5 Z^2}{c_s^4} + \frac{24 D_6 H Z^2}{c_s^4}  - \frac{20 D_7 H^2 Z^2}{c_s^4}+ \frac{108 D_8 Z^2}{c_s^6}\right)\Bigg]\nonumber  
\eea
where the $D_n$ represent all the eight different operators which appear in Eq.~(\ref{H}).\\

\end{document}